\documentclass[final,1p,times]{elsarticle}
\usepackage{amsmath}
\usepackage{amssymb}
\usepackage{epstopdf}
\newcommand{\Lambpha}{{}^{\ \ 6}_{\Lambda\Lambda}\mathrm{He}}
\newcommand{\LL}{\Lambda\Lambda}
\newcommand{\sLL}{{\scriptscriptstyle \LL}}
\newcommand{\Kmp}{K^-p}
\newcommand{\Kbn}{\bar{K}^0n}
\newcommand{\sKmp}{{\scriptscriptstyle \Kmp}}
\newcommand{\sKbn}{{\scriptscriptstyle \Kbn}}
\newcommand{\calS}{\mathcal{S}}
\newcommand{\fm}{\text{fm}}
\newcommand{\reff}{r_\text{eff}}
\usepackage[usenames]{color}
\usepackage[normalem]{ulem}  

\newcommand{\comment}[1]{}

\renewcommand\sout{\bgroup \color{red} \ULdepth=-.5ex \ULset}
\renewcommand{\sout}[1]{}

\usepackage{hyperref}
\journal{Nuclear Physics A}
\begin{document}
\rightline{YITP-16-37, KUNS-2614}

\begin{frontmatter}

\title{Hadron-Hadron Correlation and Interaction \\ from Heavy-Ion Collisions}

\author[YITP]{Akira Ohnishi}
\ead{ohnishi@yukawa.kyoto-u.ac.jp}

\author[YITP]{Kenji Morita}
\ead{kmorita@yukawa.kyoto-u.ac.jp}

\author[KyotoU]{Kenta Miyahara}
\ead{miyahara@ruby.scphys.kyoto-u.ac.jp}

\author[YITP]{Tetsuo Hyodo}
\ead{hyodo@yukawa.kyoto-u.ac.jp}


\address[YITP]{Yukawa Institute for Theoretical Physics, Kyoto University,
Kyoto 606-8502, Japan}
\address[KyotoU]{Department of Physics, Faculty of Science, Kyoto University,
Kyoto 606-8502, Japan}

\begin{abstract}
We investigate the $\Lambda\Lambda$ and $K^-p$ intensity correlations
in high-energy heavy-ion collisions.
First, we examine the dependence of the $\Lambda\Lambda$ correlation
on the $\Lambda\Lambda$ interaction
and the $\Lambda\Lambda$ pair purity probability $\lambda$.
For small $\lambda$, the correlation function needs to be suppressed
by the $\Lambda\Lambda$ interaction in order to explain
the recently measured $\Lambda\Lambda$ correlation data.
By comparison,
when we adopt the $\lambda$ value
evaluated from the experimentally measured $\Sigma^0/\Lambda$ ratio,
the correlation function needs to be enhanced by the interaction.
We demonstrate that these two cases correspond to the two analyses
which gave opposite signs of the $\Lambda\Lambda$ scattering length.
Next, we discuss the $K^-p$ correlation function.
By using the 
local $\bar{K}N$ potential
which reproduces the kaonic hydrogen data by SIDDHARTA,
we obtain the $K^-p$ correlation function.
We find that the $K^-p$ correlation can provide a complementary information
with the $K^{-}p$ elastic scattering amplitude.
\end{abstract}


\end{frontmatter}


\section{Introduction}

Interactions between hadrons are basic ingredients
in nuclear and hadron physics.
We need
nucleon-nucleon ($NN$), hyperon-nucleon ($YN$)
and hyperon-hyperon ($YY$) interactions
to theoretically investigate normal nuclear and hypernuclear structure
and nuclear matter equation of state (EOS).
$\LL$ interaction is one of the key interactions
in exotic hadron physics and neutron star physics.
First, $\LL$ interaction is closely related
to the existence of the dihyperon, the $H$ particle ($uuddss$).
While it seems improbable that there is a bound $H$ state
below the $\LL$ threshold,
$H$ may exist as a loosely bound state or as a resonance state.
The deeply bound $H$~\cite{Jaffe}
was ruled out by the observation of the double $\Lambda$ hypernucleus
$\Lambpha$ in the Nagara event~\cite{Nagara,Nagara-Update},
and the upper bound of loosely bound $H$ production
is found to be very small;
it is much lower than
the anti-deuteron ($\Upsilon(1S,2S)\to\bar{d}X$) production
at the KEKB $e^+e^-$ collider~\cite{Belle2013},
and much lower than various theoretical predictions~\cite{ExHIC}
at the Large Hadron Collider at CERN~\cite{ALICE2016}.
By comparison,
a bump structure above the $\LL$ threshold was observed~\cite{KEK-E522},
and recently performed ab-initio calculations
show the existence of the bound $H$
at least in the SU(3) limit with unphysical quark masses~\cite{LQCD,Haidenbauer:2011ah}.
The $H$ around the $\LL$ threshold should have the molecule nature
of $\LL$, and information on $\LL$ interaction is decisive.
$\LL$ interaction is also important in neutron star physics,
especially to solve the "hyperon puzzle".
Hyperons are expected to appear in the core of heavy neutron stars,
whereas the hyperonic equations of state of neutron star matter
are generally too soft to support $2 M_\odot$ neutron stars~\cite{MassiveNS}.
If the $\LL$ interaction is repulsive enough at high densities,
it may be possible to support massive neutron stars.
Contrary to its importance,
information on the $\LL$ interaction is very limited.
There is only one uniquely identified double $\Lambda$ hypernucleus,
$\Lambpha$, observed in Nagara event~\cite{Nagara,Nagara-Update}.
The bond energy $\Delta B_\sLL$ provides precious information,
but it is not enough to determine the shape of the $\LL$ potential.

Recent developments in exotic hadron physics
demand deeper understanding
of meson-baryon ($MB$) and meson-meson ($MM$) interactions.
There have been exciting developments
in the spectroscopy of hadron resonances,
starting with the discovery of 
$D_{sJ}(2317)$~\cite{DsJ2317} and $X(3872)$~\cite{X3872} in the charmed meson sector.
Recently, new states are also observed in the bottom sector, such as $Z_{b}^{\pm}(10610)$ and $Z_{b}^{\pm}(10650)$ \cite{Belle:2011aa} and in the baryon sector, $P_{c}^{+}(4380)$ and $P_{c}^{+}(4450)$~\cite{Aaij:2015tga}.
These states cannot be explained by the simple quark model
and are considered to be candidates of exotic hadrons. Among others, the hadronic molecules are closely related to the hadron-hadron interactions.
One of the typical examples of hadronic molecules is the $\Lambda(1405)$ baryon resonance which appears near the $\bar{K}N$ threshold.
It is considered as a $\bar{K}N$ quasi-bound state
in the $\bar{K}N$-$\pi\Sigma$ coupled-channel
analyses~\cite{Dalitz1960,Kaiser1995} (see Refs.~\cite{Hyodo2011,Kamiya:2016jqc} for recent reviews).
The structure of $\Lambda(1405)$ is closely related to the strength and energy dependence
of the $I=0$ $\bar{K}N$ interaction.
The uncertainty of the $\bar{K}N$ scattering amplitude at around 
the threshold is reduced by the high precision data of the kaonic hydrogen
by SIDDHARTA~\cite{SIDDHARTA} combined with the $K^{-}p$ scattering data~\cite{Ikeda:2011pi,Ikeda:2012au}. Because the low energy $K^{-}p$ scattering data was accumulated by old bubble chamber experiments with relatively large experimental uncertainties, 
new and accurate information is desired to further sharpen the description of the $\bar{K}N$ amplitude.
Precise knowledge of the $\bar{K}N$ interaction is also important to study possible bound states of $\bar{K}$ in nuclei~\cite{PL7.288,Akaishi:2002bg}.

One of the observables to get information on hadron-hadron ($hh$) interaction
is the two hadron intensity 
correlation~\cite{KooninPratt,LL82,CGreiner1989,AO2000}.
Two hadron intensity correlation is generated
mainly by quantum statistics (QS),
known as the Hanbury Brown, Twiss~\cite{HBT}
or Goldhaber, Goldhaber, Lee, Pais~\cite{GGLP} effects,
and the final state interaction~\cite{KooninPratt,LL82}.
One expects substantial dependence of the correlation function
on the pairwise $hh$ interaction,
provided that the interaction is sufficiently strong
in the range comparable to the effective source size~\cite{KooninPratt,LL82}.
Recently, $\LL$ correlation in high-energy heavy-ion collisions
has been measured 
at the Relativistic Heavy-Ion Collider (RHIC)
at the Brookhaven National Laboratory~\cite{STAR}.
Theoretical analysis of data implies that the $\LL$ interaction
is weakly attractive and there is no loosely bound state~\cite{MFO2015}.
By comparison, another analysis suggests that
$\LL$ interaction is weakly repulsive
or there is a loosely bound state~\cite{STAR}.
Resolving this contradiction is an important step
to utilize $hh$ correlation as a tool to extract $hh$ interaction.
If it is successful, 
we can apply the same method to other $hh$ correlation.
Specifically, $K^-p$ correlation is a good candidate.
Both of $K^-$ and $p$ are long-lived charged particles
and abundantly produced in heavy-ion collisions,
then it is possible to measure the correlation function precisely.
The correlation function measurement will provide a further constraint on the $\bar{K}N$ interaction, in combination with the accumulated data of the kaonic hydrogen and the $K^{-}p$ scattering.

In this article,
we investigate the $\LL$ and $K^-p$ correlations in heavy-ion collisions.
We first compare
the $\Lambda\Lambda$ correlation 
functions in the two correlation function formulae,
the Koonin-Pratt (KP) formula~\cite{KooninPratt}
and the Lednicky-Lyuboshits (LL)~\cite{LL82} model formula.
The LL model is applied
to analyze the $\LL$ correlation data in Ref.~\cite{STAR},
and the KP formula is used in Ref.~\cite{MFO2015}.
The LL model is an analytic model,
where the asymptotic wave function is assumed.
We demonstrate that these two formulae give almost
the same $\LL$ correlation function from heavy-ion collisions
for the the static and spherical source.
Next we examine the dependence of the $\Lambda\Lambda$ correlation
on the pair purity parameter $\lambda$ and the scattering length $a_0$.
The quantum statistical correlation for $\LL$ with the pair purity probability $\lambda$ reads
$C_\sLL(q\to0)=1-\lambda/2$ at small relative momentum $q$,
and the STAR data show $C_\sLL(q\to0)\simeq 0.82$.
When $\lambda$ is close to unity,
the quantum statistical correlation function approaches $C_\sLL\simeq 0.5$,
and thus the interaction should be attractive 
to explain the observed {\em enhanced} correlation.
For small $\lambda$, on the other hand,
the quantum statistical correlation function becomes close to unity
and the interaction needs to {\em suppress} the correlation.
We demonstrate that the choice of the $\lambda$ parameter
is the origin of the difference
between the results in Ref.~\cite{STAR} and Ref.~\cite{MFO2015}.
In the former,
$\lambda$ is regarded as a free parameter and the optimal value is found
to be small, then the positive scattering length $a_0>0$ 
(decreasing phase shift) is favored.
In the latter,
$\lambda$ is evaluated by using measured data~\cite{Sullivan1987,RHIC-Xi},
and the favored scattering length is found to be
$a_0<0$ (increasing phase shift).
Finally we discuss the $K^- p$ correlation function.
The $\bar{K}N$ scattering has two components, $I=0$ and $I=1$ in the isospin basis and $K^{-}p$ and $\bar{K}^{0}n$ in the charge basis. We thus write down the correlation function formula in the case of coupled-channel scattering.
To predict the correlation function, we adopt the $\bar{K}N$ potential
developed in Ref.~\cite{Miyahara2015} which reproduces the scattering amplitude ~\cite{Ikeda:2011pi,Ikeda:2012au}. We also examine the dependence of the correlation function with respect to the details of the potential.

This article is organized as follows.
In Sec.~\ref{Sec:CorrFunc},
we briefly summarize the correlation function formalism.
In Sec.~\ref{Sec:LL},
we discuss the $\LL$ correlation function
and its dependence on $\LL$ interaction.
In Sec.~\ref{Sec:Kmp},
we discuss the $\Kmp$ correlation in heavy-ion collisions.
We summarize our work in Sec.~\ref{Sec:Summary}.

\section{Hadron-Hadron Correlation Function in Heavy-Ion Collisions
and Its Relation to Interaction}\label{Sec:CorrFunc}

In this section,
we explain the correlation function formulae,
the KP formula~\cite{KooninPratt} and the LL~\cite{LL82} model formula.
The former is used in Ref.~\cite{MFO2015}
and the latter is used in Ref.~\cite{STAR}.
We compare the correlation functions in these two formulae 
in Secs.~\ref{Sec:LL} and ~\ref{Sec:Kmp}.

\subsection{Correlation Function}

Two-hadron intensity correlation from a chaotic source is given
in the KP formula~\cite{KooninPratt},
\begin{align}
C(\bold{q},\bold{P})
=&\frac{
\int d^4x_1 d^4x_2
S_1(x_1,\bold{p}_1)
S_2(x_2,\bold{p}_2)
\left| \Psi^{(-)}(\bold{r},\bold{q}) \right|^2
}{
\int d^4x_1 d^4x_2
S_1(x_1,\bold{p}_1)
S_2(x_2,\bold{p}_2)
}
\label{Eq:KP}\\
\simeq&
\int d\bold{r}
S_{12}(\bold{r})
\left| \Psi^{(-)}(\bold{r},\bold{q}) \right|^2
\ ,\label{Eq:CorrelationR}
\end{align}
where $S_i(x_i,\bold{p}_i)~(i=1,2)$ is the single particle source function
of the hadron $i$ with momentum $\bold{p}_i$,
$\bold{q}=(m_2\bold{p}_1-m_1\bold{p}_2)/(m_1+m_2)$
is the relative momentum,
$\bold{P}=\bold{p}_1+\bold{p}_2$ is the center-of-mass momentum,
$\bold{r}$ is the relative coordinate with time difference correction,
and $\Psi^{(-)}$ is the relative wave function
in the two-body outgoing state
with an asymptotic relative momentum $\bold{q}$.
In the case where we can ignore the time difference of the emission
and the momentum dependence of the source,
we integrate out the center-of-mass coordinate
and obtain Eq.~\eqref{Eq:CorrelationR},
where $S_{12}(\bold{r})$
is the normalized pair source function in the relative coordinate.

We assume here that only $s$-wave part of the wave function is modified by the hadronic interaction.
For $\LL$, the relative wave function is given as
\begin{align}
\Psi^{(-)}_{\sLL}
=\sqrt{2}\chi_s\left[ \cos (\bold{q}\cdot\bold{r}) + \psi_{\sLL}(r) - j_0(qr)\right]
+\sqrt{2}i\chi_t\sin(\bold{q}\cdot\bold{r})
\ ,
\end{align}
where $\chi_s$ and $\chi_t$ show the spin part of the wave function in spin-singlet and spin-triplet, respectively,
$j_0$ is the spherical Bessel function,
and 
$\psi_{\sLL}$ is the spatial part of the relative wave function in $s$-wave, 
which is regular at $r\to0$ and has an asymptotic form,
\begin{align}
\psi_{\sLL}(r) \to
 \frac{e^{-i\delta}}{qr}\sin(qr+\delta)
=\frac{1}{2iqr}\left[e^{iqr}-e^{-2i\delta}e^{-iqr}\right]
\quad (r\to \infty)\ ,
\label{Eq:LLwf}
\end{align}
with $\delta$ being the phase shift.
Then for the static and spherical source,
$S_i(x,\bold{p})\propto \exp(-\bold{x}^2/2R^2)\delta(t-t_0)$,
the correlation function is obtained as
\begin{align}
C^\text{sph}_{\sLL}(q)\simeq
1 - \frac12 \exp(-4q^2R^2)+\frac12 \int_0^\infty 4\pi r^2\, dr\,
S_{12}(\bold{r})
\left[
\left|\psi_{\sLL}(r)\right|^2
-\left|j_0(qr)\right|^2
\right]
\ ,\label{Eq:corr1}
\end{align}
where $S_{12}(\bold{r})=\exp(-r^2/4R^2)/(2\sqrt{\pi}R)^3$
and we take the spin average of $|\Psi^{(-)}|^2$.
The second term arises from the quantum statistical effect
which suppresses the correlation
due to the anti-symmetrization of the wave function for spin-half fermions.
The third term shows the interaction effects;
when the wave function is enhanced due to the attraction,
the correlation is enhanced accordingly.

For $K^-p$,
we consider the following wave function in the $\Kmp$ channel,
\begin{align}
\Psi^{(-)}_\sKmp
=\exp(i\bold{q}\cdot\bold{r}) + \psi_\sKmp(r) - j_0(qr)
\ ,
\end{align}
where $\psi_\sKmp$ is the $s$-wave relative wave function with the outgoing
boundary condition. The relative momentum is defined by 
$\bold{q}=(M_{p}\bold{k}_{K}-m_{K^{-}}\bold{k}_{p})/(m_{K^{-}}+M_{p})$
with $\bold{k}_{K^{-}}$ ($\bold{k}_{p}$) being the momentum of $K^{-}$ ($p$).
We ignore the Coulomb interaction in this work.
The correlation function is calculated to be
\begin{align}
C^\text{sph}_{\sKmp}(q)\simeq
1+\int_0^\infty 4\pi r^2\,dr\, S_{12}(\bold{r})
\left[
\left|\psi_{\sKmp}(r)\right|^2
-\left|j_0(qr)\right|^2
\right]
\ ,\label{Eq:corr2}
\end{align}
It should be noted that 
there is no (anti-)symmetrization of the wave function.

Since $\Kmp$ couples with $\Kbn$,
we need to take care of the channel coupling. With the isospin basis wave function $\psi_{I}(r)$ which has the asymptotic form $\sin(qr+\delta_I)/(qr)$, a general form of the $\bar{K}N$ wave function $\Psi^{(-)}_{{\scriptscriptstyle \bar{K}N},\ell=0}$ can be written as the superposition of the isospin components
\begin{align}
\Psi^{(-)}_{{\scriptscriptstyle \bar{K}N},\ell=0}
&=C_{0}\frac{\chi(\Kmp)+\chi(\Kbn)}{\sqrt{2}}\psi_0(r)
+C_{1}\frac{-\chi(\Kmp)+\chi(\Kbn)}{\sqrt{2}}\psi_1(r)
\ , \\
&=\chi(\Kmp)\psi_\sKmp(r)
+\chi(\Kbn)\psi_\sKbn(r)
,
\end{align}
where $\chi(\Kmp)$ and $\chi(\Kbn)$ denote the isospin wave function specifying the charge state\footnote{The phase convention is chosen to be $|K^{-}\rangle=-|I=1/2,I_{3}=-1/2\rangle$.}.
In order to keep the outgoing boundary condition with purely $\Kmp$ channel,
the coefficients $C_{0}$ and $C_{1}$ should be
\begin{align}
C_{0} = & \frac{e^{-i\delta_0}}{\sqrt{2}}
\ ,\quad
C_{1} = -\frac{e^{-i\delta_1}}{\sqrt{2}}
\ ,
\end{align}
where $\delta_I (I=0,1)$ shows the phase shift in the isospin base.
The outgoing wave in the $\Kbn$ channels disappears,
and the asymptotic wave function in the $\Kmp$ channel is found to be
\begin{align}
\psi_{\sKmp}(r)
\to&\frac{1}{2iqr}\left[e^{iqr}-\tilde{\calS}^{-1}_\sKmp e^{-iqr}\right]
\ ,\quad
\tilde{\calS}_\sKmp=2\left(\calS_{0}^{-1}+\calS_{1}^{-1}\right)^{-1} ,\quad
\calS_{I}=e^{2i\delta_{I}}.
\label{Eq:wfKmp}
\end{align}
It should be noted that $\tilde{\calS}_\sKmp$ does not correspond to the $S$-matrix in the $\Kmp$ channel $\calS_\sKmp=(\calS_{0}+\calS_{1})/2$.
The $\bar{K}N$ channel couples with the lower threshold $\pi\Sigma$ channels,
then the phase shifts become complex, $\delta_0, \delta_1 \in \mathbb{C}$.

\subsection{Lednicky and Lyuboshits Model}

In order to examine the interaction dependence of the correlation function,
an analytic model developed by Lednicky and Lyuboshits (LL)~\cite{LL82}
is useful.
In the LL model, 
the correlation function is obtained by using the asymptotic wave function
and the effective range correction,
then it is given in terms of the scattering amplitude and the effective range.
We consider here the following asymptotic wave function,
\begin{align}
\psi_\text{asy}(r)=
\calS^{-1}\left[
\frac{\sin{qr}}{qr}+f(q)\frac{e^{iqr}}{r}
\right]
\ ,\label{Eq:LLwf3}
\end{align}
where $f(q)=(\calS-1)/2iq$.
We can evaluate the integral in the KP formula for $\psi_\text{asy}$ as
\begin{align}
\int_0^\infty dr\,S_{12}(r) |\psi_\text{asy}(r)|^2
=
\frac{1}{|\calS|^2}
\left[
\frac{|f(q)|^2}{2R^2}
+\frac{2\text{Re}f(q)}{\sqrt{\pi}R}\,F_1(x)
-\frac{\text{Im}f(q)}{R}\,F_2(x)
+\frac{F_2(x)}{x}
\right]
\ ,
\label{Eq:LLeq1}
\end{align}
where 
$x=2qR$,
$F_1(x)=\int_0^x dt e^{t^2-x^2}/x$ and $F_2(x)=(1-e^{-x^2})/x$,
and we have utilized the relation
$\int_0^\infty dt e^{-t^2+2ixt}=\sqrt{\pi}e^{-x^2}/2+ixF_1(x)$.
We find that the function $F_1$ is well approximated in the form
$F_1(x)=(1+c_1x^2+c_2x^4+c_3x^6)/(1+(c_1+2/3)x^2+c_4x^4+c_5x^6+c_3x^8)$
with $(c_1,\cdots,c_5)=(0.123,0.0376,0.0107,0.304,0.0617)$
in the $x$ range of interest, $0<x<20$~\cite{Hyp2015-Ohnishi}.
In the single channel case,
the deviation from the asymptotic wave function at small $q$ can be obtained
by using the effective range formula~\cite{RoyNigam},
\begin{align}
\lim_{q \to 0} \frac{1}{|f(q)|^2}\,\int_0^\infty r^2 dr \left[
|\psi|^2 - \frac{\sin^2(qr+\delta)}{q^2r^2} \right]
=-\frac12\,r_\text{eff}
\ .
\label{Eq:LLeq2}
\end{align}
The integral in the left hand side of Eq.~\eqref{Eq:LLeq2}
gives the correction to Eq.~\eqref{Eq:LLeq1},
when we multiply a factor $e^{-r^2/4R^2}$ to the integrand.
We expect that the factor does not change the integral much
as long as $R$ is large enough compared with the interaction range.
Under this assumption,
we can evaluate the effective range correction.
By using Eqs.~\eqref{Eq:LLeq1} and \eqref{Eq:LLeq2},
one arrives
at the interaction dependent part of the correlation function
in the LL model~\cite{LL82},
\begin{align}
\Delta C^\text{LL}(q)
=&\frac{1}{|\calS|^2}\left[
 \frac{|f(q)|^2}{2R^2}\,F_3\left(\frac{r_\text{eff}}{R}\right)
+\frac{2\text{Re}f(q)}{\sqrt{\pi}R} F_1(x)
-\frac{\text{Im}f(q)}{R} F_2(x)\right]
+\frac{1-|\calS|^2}{|\calS|^2}\frac{F_2(x)}{x}
\ ,\label{Eq:LL}
\end{align}
where $x=2qR$ and the effective range correction appears in 
$F_3(\reff/R)=1-\reff/2\sqrt{\pi}R$.
In the formula given in Ref.~\cite{LL82},
one assumes $\psi^{(-)}=(\psi^{(+)})^*$ and $|\calS|=1$,
then the $f$ is the actual scattering amplitude
and the last term in Eq.~\eqref{Eq:LL} does not exist.

\begin{figure}[tbh]
\begin{center}
\includegraphics[width=12cm]{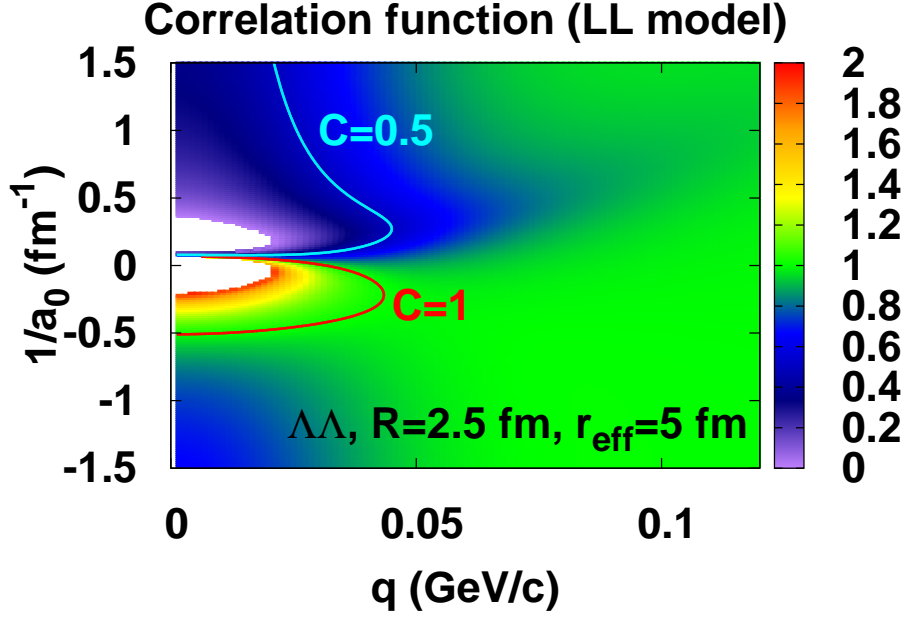}%
\end{center}
\caption{Correlation function $C_{\LL}$ as a function of $q$ and $1/a_0$
in the LL model~\cite{LL82}.
We show the results for $R=2.5~\text{fm}$ and $r_\text{eff}=5~\text{fm}$
as an example.
In the white areas, the correlation function is greater than 2
or less than 0.
}
\label{Fig:3DCLL}
\end{figure}

The $\LL$ correlation function in the LL model is given as
\begin{align}
C^\text{LL}_\sLL(q)=&1 - \frac12 e^{-4R^2q^2}
+\frac12 \Delta{C}^\text{LL}(q)
\ ,\label{Eq:LL-LL}
\end{align}
We note that there is no open channel and $|\calS|=1$ for $\LL$ at small $q$,
then the last term in Eq.~\eqref{Eq:LL} disappears.
In Fig.~\ref{Fig:3DCLL},
we show the $\LL$ correlation function
in the LL model
as a function of the relative momentum $q$
and the reciprocal of the scattering length $1/a_0$
for $R=2.5~\text{fm}$ and $r_\text{eff}=5~\text{fm}$.
It should be noted that we take the ``nuclear physics'' convention
for the scattering length, 
$q\cot\delta=-1/a_0 + r_\text{eff}q^2/2+\mathcal{O}(q^4)$,
which leads to $\delta \simeq -a_0 q$ at low energy.
When $|a_0|$ is small, the correlation function is approximately described
by the quantum statistics term, and converges to 0.5 at $q\to 0$.
In the negative $a_0$ case (attractive potential without loosely bound states),
the correlation function is enhanced especially at small $q$,
because of the enhanced wave function by the attraction.
We note that the correlation is generally suppressed
when the scattering length is positive;
Positive $a_0$ means that there is a shallow bound state
or the interaction is repulsive,
then the squared wave function is suppressed by the node or by the repulsion.
Thus the correlation function is sensitive to the $\LL$ interaction,
as long as other effects do not wash out the above trend.

As for the $K^-p$ correlation function in the LL model,
we use the wave function in the $\Kmp$ channel, Eq.~\eqref{Eq:wfKmp}, 
and find the correlation function is obtained as
\begin{align}
C^\text{LL}_\sKmp(q)=&1
+\Delta{C}^\text{LL}(q,\,\calS\to\tilde{\calS}_\sKmp,\,
f\to\tilde{f},
F_3 \to 1)
\ ,\label{Eq:Kmp-LL}
\end{align}
where $\Delta C^\text{LL}(q)$ is given in Eq.~\eqref{Eq:LL} with $F_3=1$.
It should be noted again that the ``scattering amplitude''
$\tilde{f}=(\tilde{\calS}_\sKmp-1)/2iq$
used in $\Delta C^\text{LL}$ 
is not the scattering amplitude in $\Kmp \to \Kmp$ scattering, $f_\sKmp$.
The scattering amplitude of $\Kmp$ is given as
$f_\sKmp =(f_0+f_1)/2$, where the the isospin base scattering amplitudes
are given as $\calS_I = e^{2i\delta_I} = 1+2iq f_I$.
Only when the scattering amplitudes are small $q|f_I| \ll 1$,
$\tilde{f}$ approximately matches with $f_\sKmp$.

\section{$\LL$ Correlation}\label{Sec:LL}

The $\LL$ correlation from heavy-ion collisions
has been expected to provide information
on the $\LL$ interaction~\cite{CGreiner1989,AO2000},
and it is recently measured at RHIC by the STAR collaboration~\cite{STAR}.
One of the theoretical analyses of data
implies that the scattering length of the $\LL$ interaction is negative
(increasing phase shift at low energy),
$-1.25~\fm < a_0 < 0$~\cite{MFO2015},
while the STAR collaboration concluded that the scattering length
is positive (decreasing phase shift),
$a_0=1.10\pm 0.37^{+0.08}_{-0.68}~\text{fm}$~\cite{STAR}.
The positive scattering length suggests that
there is a bound state of $\LL$ or the $\LL$ interaction is repulsive,
neither of which are not immediately acceptable,
then we now discuss the reason of the difference.

There are three differences in these analyses;
the correlation function formula, the source function,
and the assumption on the pair purity probability.
In this section,
we re-analyze the data by using the LL model
with different assumptions on $\lambda$
in order to pin down the origin of the difference in the scattering length
of the $\LL$ interaction.
We first compare the $\LL$ correlation 
in the KP and LL formulae in Subsec.~\ref{Subsec:LL1}.
We also discuss the collective flow effects.
Next we discuss the feed-down effects and the residual correlation
in Subsec.~\ref{Subsec:LL2}.
We emphasize that
the pair purity parameter $\lambda$ is the key quantity.
In Subsec.~\ref{Subsec:LL3}, 
we discuss the favored $\LL$ interactions.
Throughout this paper,
we use the minimum bias data ($0-80 \%$ centrality)
of combined $\LL$ and $\bar{\Lambda}\bar{\Lambda}$ correlation
from Au+Au collisions
at $\sqrt{s_{{\scriptscriptstyle NN}}}=200~\mathrm{GeV}$~\cite{STAR}.

\subsection{$\LL$ Correlation
in the Koonin-Pratt and Lednicky-Lyuboshits formula}\label{Subsec:LL1}

We first discuss the difference coming from the correlation function
formula.
The KP formula given in Eq.~\eqref{Eq:KP} is used in Ref.~\cite{MFO2015},
and the LL formula in Eq.~\eqref{Eq:LL-LL} is adopted in Ref.~\cite{STAR}.
We need explicit potentials to evaluate the wave function in the KP formula.
Most of the $\LL$ potentials examined in Ref.~\cite{MFO2015}
do not predict the existence of the $\LL$ bound state,
and the positive $a_0$ region is not well explored. 
Thus we re-analyze the data by using the LL model.

\begin{figure}[tbh]
\begin{center}
\includegraphics[width=7cm]{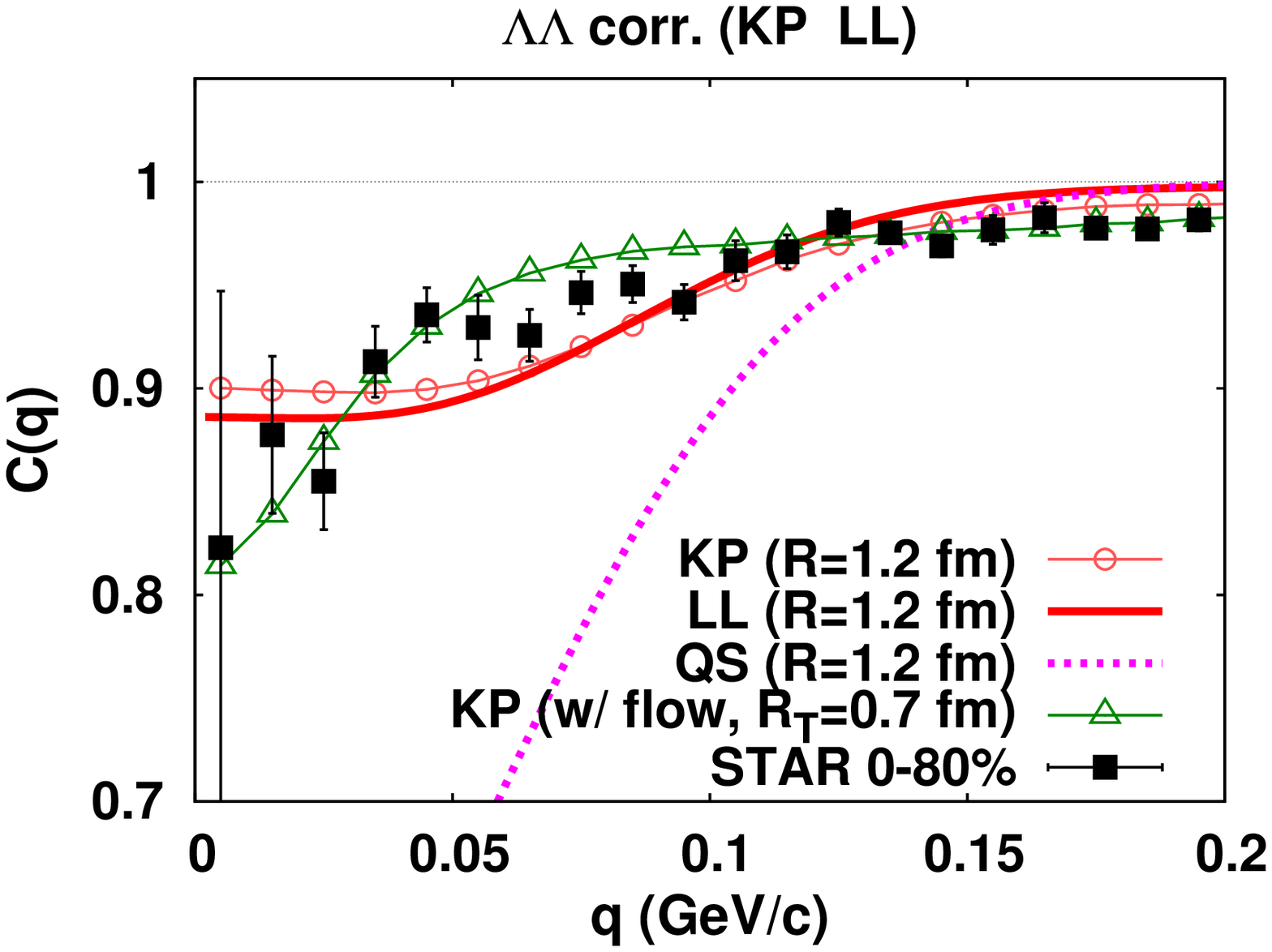}\\
\includegraphics[width=7cm]{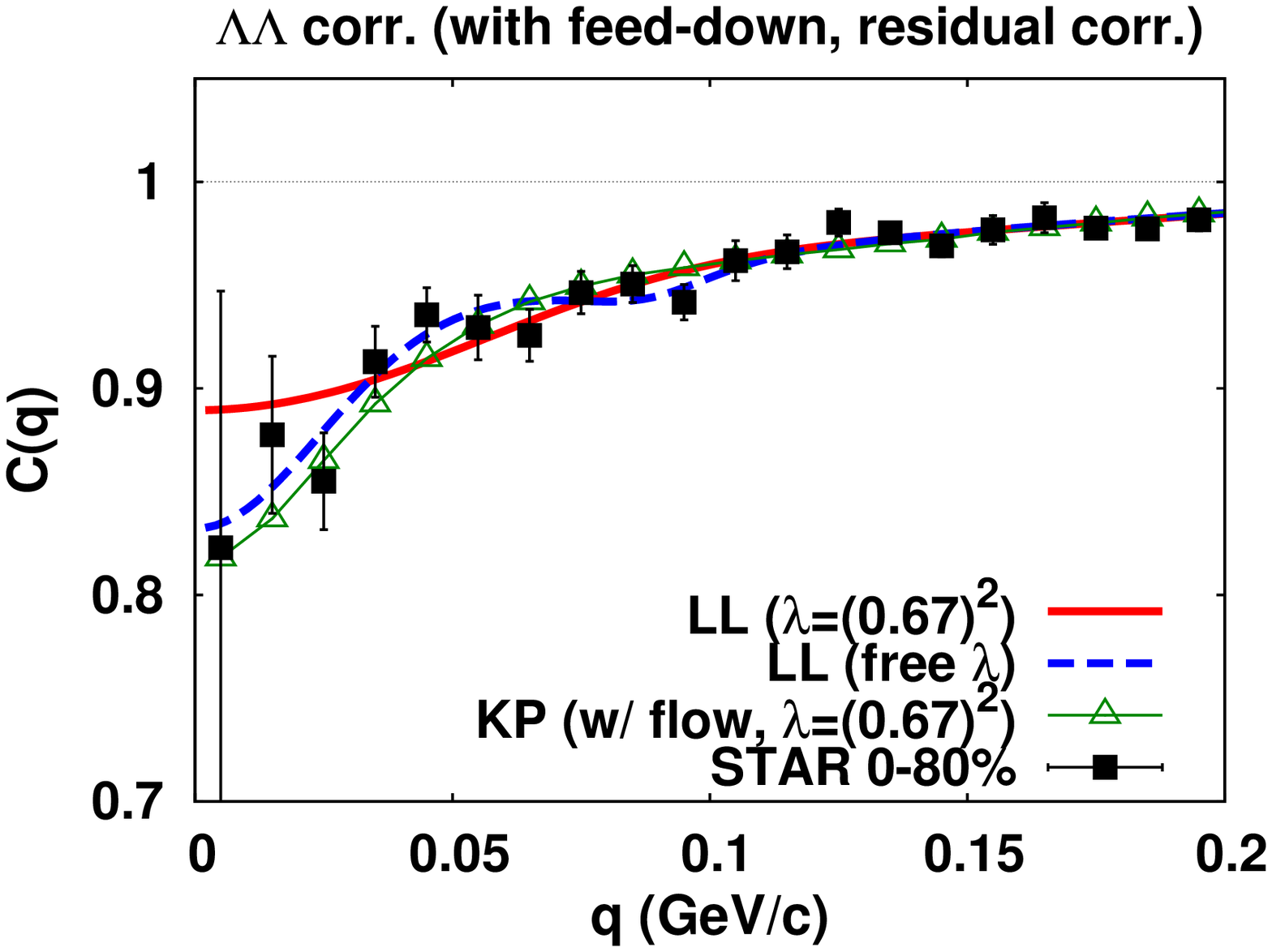}
\end{center}
\caption{$\Lambda\Lambda$ correlation function
obtained by using the KP and LL formulae
in comparison with data~\cite{STAR}.
Upper panel shows the results
with the fss2 $\LL$ potential~\cite{fss2}
from the static spherical source in the KP and LL formulae.
Lower panel shows the results
with the feed-down and residual source effects in the LL formula.
We compare the results in the fixed $\lambda$ case ($\lambda=(0.67)^2$)
and the free $\lambda$ case.
In both panels, we also show the results 
including flow effects in the KP formula~\cite{MFO2015}.
}\label{Fig:CLL}
\end{figure}

In the upper panel of Fig.~\ref{Fig:CLL},
we show the correlation function obtained
with fss2 $\LL$ interaction~\cite{fss2}
and a static spherical source, as an example.
We compare the results
in the KP (red squares) and LL formula (orange dashed line).
The optimal source size is found to be $R=1.2~\fm$ in the analysis
using the KP formula.
Because of the attraction, the correlation function in the KP formula
is enhanced
from the free results (dotted line), and approximately explains the data.
In the LL model, we take the low energy scattering parameters of fss2,
$(a_0,r_\text{eff})=(-0.81~\text{fm}, 3.99~\text{fm})$,
and calculate the phase shift from these values
with the same source size, $R=1.2~\fm$.
We find that the LL model well reproduces the correlation
in the KP formula,
suggesting that results from the asymptotic wave function corrected
with the effective range give a good estimate, even
in the case where the source size is smaller than the effective range. 
Thus the correlation functions in the KP and LL formulae are
found to be consistent for $\LL$ correlation from heavy-ion collisions,
as long as we adopt a spherical static source.

We can take account of the collective flow effects 
by modifying the source function in the KP formula.
The boost-invariant (Bjorken) expansion is assumed for the longitudinal flow,
and the transverse flow strength is fixed by fitting the transverse 
momentum spectrum of $\Lambda$.
The optimal transverse source size is found
to be $R_T\simeq 0.7~\fm$~\cite{MFO2015}.
We show the results with flow effects in the upper panel of Fig.~\ref{Fig:CLL}
(open circles).
The source is effectively elongated in the longitudinal direction,
then the correlation appears more strongly in the small $q$ region
albeit with the smaller transverse size.
This captures the feature found in the data.

\subsection{Feed-down and residual correlation effects}\label{Subsec:LL2}

The above results are not yet satisfactory in two points.
First, we have not taken account of $\Lambda$ emission
from the decay of long-lived particles.
The discussion so far applies to the case where $\Lambda$ particles are
directly emitted from the hot matter.
Feed-down from short-lived hyperon resonances can be taken into account
by modifying the source size,
and weak decay from $\Xi$ and $\Omega$ can be rejected
by using the distance of closest approach to the primary vertex~\cite{STAR}.
By contrast, we cannot reject $\Lambda$ from $\Sigma^0$,
which decays electromagnetically.
Second, the optimal transverse source size is much smaller
than that expected from the correlation analyses of other hadrons.
In the STAR data, we find that the $\LL$ correlation function is suppressed
significantly even at high relative momentum region, $q \sim 0.2~\text{GeV}$.
This high-momentum tail may suggest the existence of
unknown smaller-size source, referred
to as the "residual" source~\cite{STAR},
and makes the favored source size smaller.
One can take
account of the feed-down effects and the residual source effects
by modifying the correlation function as follows,
\begin{align}
C_\text{corr}(q)=\mathcal{N}\left(1+\lambda(C_\text{bare}(q)-1)
+a_\text{res}e^{-4r_\text{res}^2q^2}\right)
\ ,
\label{Eq:FeedRes}
\end{align}
where $C_\text{bare}(q)$ is given in the KP or LL formula,
Eq.~\eqref{Eq:KP} or Eq.~\eqref{Eq:LL-LL}, and $\mathcal{N}$ shows
the global normalization factor.
The pair purity probability $\lambda$ receives an apparent reduction
when significant part of $\Lambda$ comes from $\Sigma^0\to\gamma\Lambda$.
As a result, the deviation from unity ($C-1$) is suppressed.
The last term represents the modification by the residual source.

One of the differences in the two analyses~\cite{MFO2015,STAR}
is the assumption on the pair purity probability $\lambda$.
In Ref.~\cite{MFO2015}, we
have evaluated $\lambda$ based on the measurements
of $\Sigma^0$ and $\Xi$ (fixed $\lambda$ case),
while the STAR collaboration takes $\lambda$ as a free parameter
(free $\lambda$ case).
The pair purity
probability $\lambda$ may be evaluated as
$\lambda=(0.67)^2=((1-0.278-0.15)/(1-0.15))^2=0.4489$~\cite{MFO2015} 
based on the observed ratio
$\Sigma^0/\Lambda_\text{tot}=0.278$~\cite{Sullivan1987}
and $(\Xi\to\Lambda)/\Lambda_\text{tot}=0.15$~\cite{RHIC-Xi},
where $\Lambda_\text{tot}$ represents $\Lambda$ yield including
decay contributions.
While the above $\Sigma^0/\Lambda$ ratio is measured
in a different reaction, it is close to the statistical model estimate
and small modification of $\lambda$ does not change our conclusion.
Readers may doubt that the above pair purity probability $\lambda$
is too large compared with the measured pair purity probability
in the $p\Lambda$ correlation, $\lambda \simeq 0.15$~\cite{STARpL}.
It should be noted that, however, protons and $\Lambda$s in Ref.~\cite{STARpL}
include those from weak as well as electromagnetic decays.
They also include misidentified protons and $\Lambda$s
from the energy loss and combinatorial background, respectively.
By comparison, $\Lambda$s are identified by the weak decay vertex
in Ref.~\cite{STAR}, then we can ignore the combinatorial background.
By using the identification efficiency ($86 \pm 6 \%$) 
and the evaluated primary fraction ($45 \pm 4 \%$) for $\Lambda$s
in Ref.~\cite{STARpL} and the $\Xi$ weak decay contribution 
(15 \%)~\cite{RHIC-Xi},
the relevant purity of $\Lambda$s may be evaluated as
$\Lambda/(\Lambda+\Sigma^0)\simeq 0.45/0.86/(1-0.15) \simeq 0.62$.
This value is a little smaller but is consistent with the estimate 
$\Lambda/(\Lambda+\Sigma^0)\simeq 0.67$ in Ref.~\cite{MFO2015}
within the range of error.

\begin{table}
\caption{Optimized parameters for the $\LL$ correlation
in the fixed and free $\lambda$ cases in the LL model.
Numbers in the parentheses for $\chi^2/\mathrm{DOF}$ and DOF
show those for a given $(1/a_0,\reff)$.
In the fixed $\lambda$ case,
$1/a_0$ and $\reff$ are strongly correlated with $a_\mathrm{res}$.
Errors in the brackets in the fixed $\lambda$ case are those
in the fixed $a_\mathrm{res}$ case.
}
\label{Tab:pars}
\centerline{
\begin{tabular}{lcc}
\hline
\hline
		& Fixed $\lambda$ case		& Free $\lambda$ case	\\
\hline
$\lambda$	& $(0.67)^2=0.4489$ 		& $0.18 \pm 0.05$	\\
$1/a_0\ (\fm^{-1})$
		& $-1.26\pm 0.74\ [\pm 0.17]$	& $0.91 \pm 0.20$	\\
$\reff\ (\fm)$
		& $1.76\pm 11.62\ [\pm 0.86]$	& $8.51 \pm 2.14$	\\
$R\ (\fm)$	& $1.39\pm 0.71\  [\pm 0.17]$	& $2.88 \pm 0.38$	\\
$r_\mathrm{res}\ (\fm)$
		& $0.48\pm 0.10\  [\pm 0.02]$	& $0.43 \pm 0.03$	\\
$a_\mathrm{res}\ (\fm)$
		& $-0.058\pm 0.069$ [fixed]	& $-0.045 \pm 0.004$	\\
$\mathcal{N}$	& $1.006\pm 0.001\ [\pm 0.001]$	& $1.006 \pm 0.001$	\\
\hline
$\chi^2/\mathrm{DOF}$
		& $0.64\ (0.61)\ [0.63]$	& $0.55 (0.53)$		\\
DOF		& $44\ (46)\ [45]$		& $43 (45)$		\\
\hline
\hline
\end{tabular}}
\end{table}

In the lower panel of Fig.~\ref{Fig:CLL},
we compare the results
in the fixed $\lambda$ (solid line) and free $\lambda$ (dashed line) cases
in the LL model,
where the best fit parameters 
and $\chi^2/\mathrm{DOF}$ are obtained as
$(a_0, \reff, R, \chi^2/\text{DOF})=(-0.79~\fm, 1.8~\fm, 1.4~\fm, 0.64)$
and 
$(1.10~\fm, 8.5~\fm, 2.9~\fm, 0.55)$,
respectively.
Other parameters and errors are summarized in Table~\ref{Tab:pars}.
We have confirmed that positive $a_0$ values are favored 
in the free $\lambda$ case
and the optimal value is found to be $\lambda\simeq 0.18$,
which is consistent with the STAR collaboration result~\cite{STAR}.
Quantum statistics and the pair purity give
$C(q\to0)=1-\lambda/2 \sim 0.91$ at $\lambda=0.18$,
while the data show $C(q\to0) \simeq 0.82$.
Thus we need to suppress $C(q)$ at small $q$ by the $\LL$ interaction
and positive $a_0$ is favored.
By contrast, for a fixed $\lambda=(0.67)^2$,
the corresponding quantum statistical correlation
$C_\sLL(q\to0)=1-\lambda/2\simeq 0.78$
is slightly smaller than the observed correlation.
With the residual source contribution, $a_\text{res}\sim -0.06~\fm$,
the difference from the data becomes more evident.
The $\LL$ interaction needs 
to enhance the correlation,
and the optimal $a_0$ value is found in the negative region,
as concluded in Ref.~\cite{MFO2015}.

Flow effects may be also important for quantitative discussions.
In the fixed $\lambda$ case,
the correlation function in the LL model
overestimates the data at small $q$,
which may be improved when we take account of the flow effects.
For example, the results in the KP formula
with flow and the fss2 $\LL$ potential effects
show suppression at small $q$~\cite{MFO2015}
as shown by triangles in the lower panel of Fig.~\ref{Fig:CLL},
which give a better description at small $q$.

\subsection{Favored $\LL$ interactions}\label{Subsec:LL3}

\begin{figure}[tbh]
\begin{center}
\includegraphics[width=12cm]{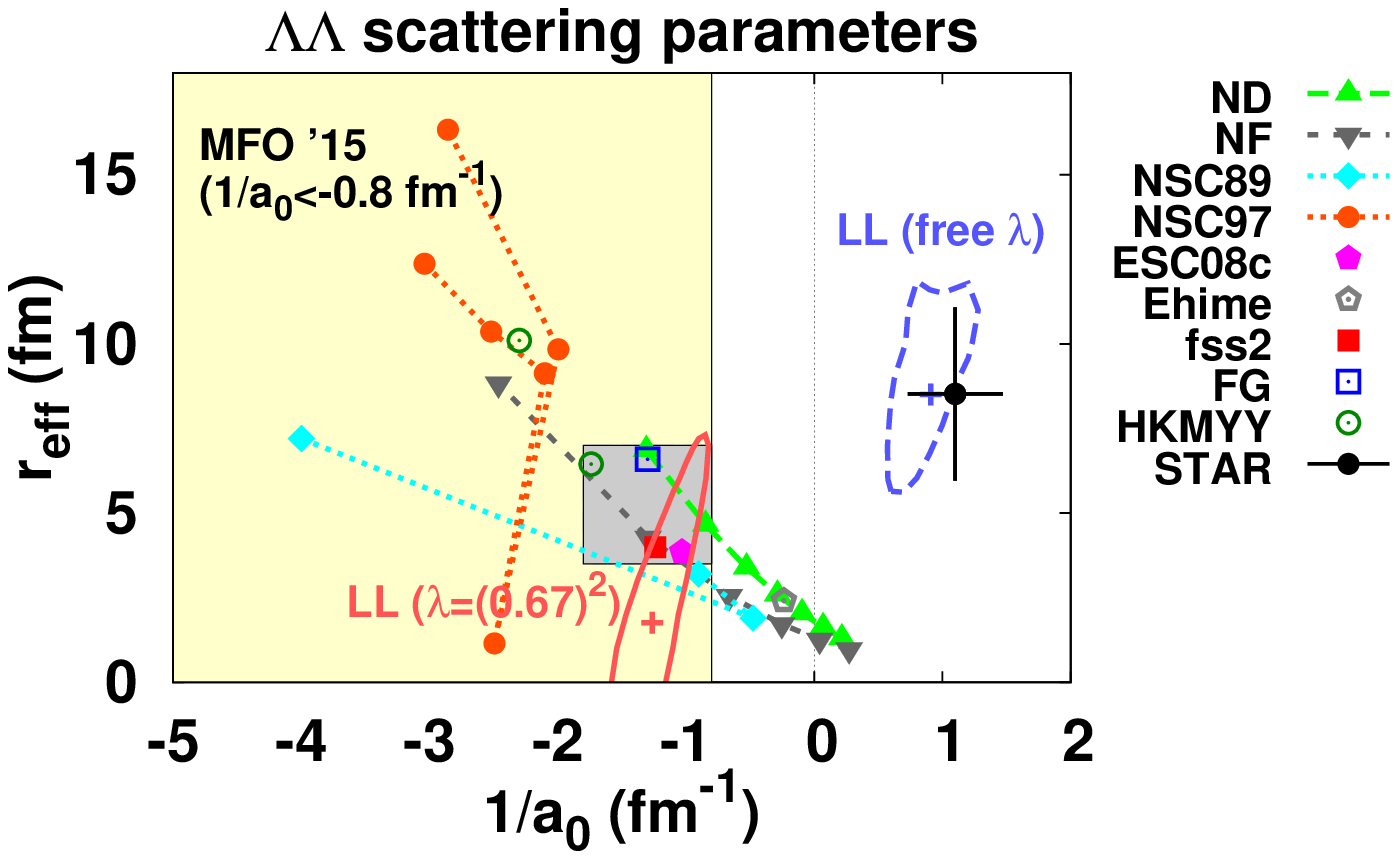}\\[2ex]
\end{center}
\caption{
Low-energy scattering parameters $(a_0,r_\text{eff})$ of $\LL$.
Contours show $\chi^2/\text{DOF}=0.65$ ($\lambda=(0.67)^2$, solid contour)
and $\chi^2/\text{DOF}=0.56$ (free $\lambda$, dashed contour)
in the LL model analysis of the $\LL$ correlation data.
Symbols show $(1/a_0,r_\text{eff})$ from $\LL$
potentials~\cite{fss2,NHC,NSC,ESC08,Ehime,FG,HKMYY,HKYM2010},
and shaded areas show the region favored by the $\LL$ correlation data
in Ref.~\cite{MFO2015}(MFO '15).
Filled black circle with $xy$ error bar shows
the analysis result by the STAR collaboration,
where $\lambda$ is regarded as a free parameter~\cite{STAR}.
}
\label{Fig:ar}
\end{figure}

In Fig.~\ref{Fig:ar}, we show
the favored region boundary $\chi^2/\text{DOF}=0.65$ ($0.56$)
in the fixed (free) $\lambda$ case in the LL model.
The region in the free $\lambda$ case is consistent with that
by the STAR collaboration~\cite{STAR}.
As in the best fit results shown in the previous subsection,
negative and positive scattering lengths are favored
in the fixed and free $\lambda$ cases, respectively.
We find that negative scattering lengths are more favored
in the pair purity probability range of $\lambda > 0.35$;
The $\chi^2/\mathrm{DOF}$ at the negative $a_0$ local minima
is smaller than that at the positive $a_0$ local minima
when $\lambda$ is fixed at a value $\lambda > 0.35$.

In Fig.~\ref{Fig:ar}, we also show
the low energy scattering parameters $(1/a_0,r_\text{eff})$ of
several $\LL$ interactions;
Boson exchange potentials 
(ND, NF, NSC89, NSC97, ESC08, Ehime)~\cite{NHC,NSC,ESC08, Ehime}
and Nijmegen-based potentials
fitted to the Nagara data (FG,HKMYY)~\cite{FG,HKMYY,HKYM2010},
in addition to the quark model potential (fss2)~\cite{fss2}.
We note that the fixed $\lambda$ region covers
recently proposed $\LL$ potentials, fss2 and ESC08~\cite{fss2,ESC08}.

The shared areas in Fig.~\ref{Fig:ar}
show the favored region in the analysis using the KP formula~\cite{MFO2015}.
The dark (grey) shaded area shows the region with $\chi^2/\mathrm{DOF}<5$
with flow effects but without feed-down and residual correlation effects.
The light (yellow) shaded area shows the region with $\chi^2/\mathrm{DOF} \lesssim 1$
under the condition $R > r_\mathrm{res}$
with flow, feed-down and residual correlation effects.
We note that the light shaded area includes
the favored region in the fixed $\lambda$ case in the LL model analysis.

\section{$K^-p$ Correlation}
\label{Sec:Kmp}

The $\bar{K}N$ interaction is the key to understand the structure of the $\Lambda(1405)$ and the properties of $\bar{K}$ in nuclear medium. There have been a long-standing problem of the inconsistency between the $K^{-}p$ scattering data and the kaonic hydrogen measurement. The problem has eventually been resolved by the new result of the kaonic hydrogen from the SIDDHARTA collaboration~\cite{SIDDHARTA}. Thanks to the precise measurement by SIDDHARTA, quantitative understanding of whole experimental database is now achieved by the coupled-channel approach with chiral SU(3) dynamics 
at the level of $\chi^{2}/\text{DOF}\sim 1$~\cite{Ikeda:2011pi,Ikeda:2012au}. To predict the correlation function, we need the $K^{-}p$ wave function $\psi_{\sKmp}(r)$, which can be calculated by the equivalent local potential as developed in Ref.~\cite{HyodoWeise}. The most reliable $\bar{K}N$ potential at present is constructed in Ref.~\cite{Miyahara2015}, using the scattering amplitude of Refs.~\cite{Ikeda:2011pi,Ikeda:2012au}. By construction, this $\bar{K}N$ potential reproduces the experimental data with the accuracy of $\chi^{2}/\text{DOF}\sim 1$.

\begin{figure}[bthp]
\begin{center}
\includegraphics[width=8cm]{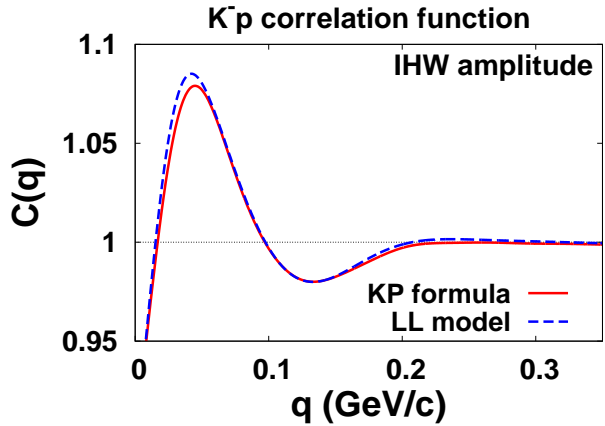}
\end{center}
\caption{$K^{-}p$ correlation function obtained by Eq.~\eqref{Eq:corr2} with the potential in Ref.~\cite{Miyahara2015} based on the NLO chiral SU(3) dynamics~\cite{Ikeda:2011pi,Ikeda:2012au} (solid line) and that obtained by the LL model formula~\eqref{Eq:Kmp-LL} with the same amplitude (dashed line). The source size is set to be $R=3~\text{fm}$.
}
\label{Fig:Kmp}
\end{figure}

In Fig.~\ref{Fig:Kmp}, we show the $K^{-}p$ correlation function obtained by the $\bar{K}N$ potential in Ref.~\cite{Miyahara2015}.
The source size of nonidentical particle pairs 
can be estimated as $R=\sqrt{(R_K^2+R_p^2)/2}$.
The kaon source size in Au+Au collisions at
$\sqrt{s_{{\scriptscriptstyle NN}}}=200~\mathrm{GeV}$
is estimated as $R_K=(2-5)~\fm$~\cite{PHENIXkaon,STARkaon},
and the proton source size is expected to be similar.
We here take $R=3.0$ fm.
The interaction range of the original potential is about 0.4 fm~\cite{Miyahara2015}. To examine the potential range dependence, we prepare different potentials by changing the range parameter but
keeping the amplitude unchanged. We find that the correlation function does not change very much when we vary the range parameter from 0.2 fm to 0.8 fm.
Because there is no $\pi$ exchange in the $K^{-}p$ system, we conclude that the short range details of the $K^{-}p$ interaction does not affect the correlation function for the source size $R=3.0$ fm.
Thus the correlation function is dominated by the long range part 
of the wave function, and the correlation function is well reproduced
by the LL model~\eqref{Eq:Kmp-LL}, as shown by the dashed line in Fig.~\ref{Fig:Kmp}.
We note that the Coulomb interaction is not included in the present result. The inclusion of the Coulomb interaction will modify the correlation function at small $q$. In the actual measurement, there is the $\Lambda(1520)$ resonance in $d$-wave $K^{-}p$ scattering, which may affect the correlation around $q\sim 0.24$ GeV/c.

It is also interesting to note the bump structure around $q\sim 0.05$ GeV/$c$.
There is no bump structure in the $K^{-}p$ amplitude at the corresponding energy. It turns out that this bump structure arises from the detailed interference between two phases of $I=0$ and $I=1$ components in $\tilde{S}_{\sKmp}$ defined in Eq.~\eqref{Eq:wfKmp}. In this way, the $K^{-}p$ correlation function gives a complementary information with the elastic $K^{-}p$ scattering.

\section{Summary}
\label{Sec:Summary}

We have analyzed the $\LL$ and $\Kmp$ intensity correlation
in high-energy heavy-ion collisions,
which will provide information on the $\LL$ and $\Kmp$ interactions.

We have investigated the dependence of the $\LL$ correlation
on the $\LL$ interaction and the pair purity parameter $\lambda$.
Recent two analyses of the $\LL$ correlation data~\cite{STAR,MFO2015}
give different signs of the scattering length for the favored $\LL$ interaction.
This difference is found to come from
the assumption on the pair purity parameter $\lambda$.
When $\lambda$ is chosen to minimize the $\chi^2$,
the optimal value of $\lambda$ is found to be small, $\lambda\simeq 0.18$.
The corresponding quantum statistical
correlation is larger than the observed value, $C_\sLL(q\to0)\simeq 0.82$,
then the $\LL$ interactions with positive $a_0$
(decreasing phase shift at low energy) are favored
in order to suppress the correlation.
With $\lambda=(0.67)^2$ evaluated on the basis of the measured 
data of the $\Sigma^0/\Lambda$~\cite{Sullivan1987,RHIC-Xi},
the corresponding quantum statistical correlation
is smaller than the observed correlation.
Thus the $\LL$ interactions with negative $a_0$ (increasing phase shift)
are favored to enhance the correlation.
Experimental confirmation of $\Sigma^0$ yield in heavy-ion collisions
is important.

We have also discussed the $\Kmp$ correlation function
in heavy-ion collisions. We use the $\Kmp$ potential developed in Ref.~\cite{Miyahara2015}
which is fitted to the scattering amplitude
including the SIDDHARTA data~\cite{Ikeda:2011pi,Ikeda:2012au}.
We find that the $K^{-}p$ correlation function does not depend on the short range details of the potential very much, for the source size of $R\sim 3$ fm.
Because of the coupled-channel nature of the problem, the $K^{-}p$ correlation function reflects a particular combination of the isospin components which is different from the $K^{-}p$ elastic scattering. This is a unique feature of correlation functions in coupled channel systems. As a consequence, the detailed study of the $K^{-}p$ correlation function is considered as complementary to the $K^{-}p$ scattering.

There are more works to be done as an extension of this work.
As for the $\LL$ interaction,
comparison with data obtained 
at the KEKB $e^+e^-$ collider~\cite{Belle2013}
as well as data to be obtained 
at the Large Hadron Collider at CERN~\cite{ALICE2016}
and J-PARC~\cite{J-PARC_E42} should 
be helpful to constrain $\LL$ interaction more precisely.
Understanding the origin of the  ``residual'' source is a theoretical challenge.
It is also important to utilize the dynamical model source function.
As for the $\Kmp$ correlation,
the Coulomb interaction has to be seriously considered,
since it will modify the correlation at small $q$.
Application to other channels such as $\Omega^-p$~\cite{Hyp2015-Morita}
is also interesting.

\section*{Acknowledgement}
Numerical computations were carried out on SR16000 at YITP in Kyoto university.
This work is supported in part 
by the Grants-in-Aid for Scientific Research from JSPS
(Nos. 15K05079,  
      15H03663, 
and   24740152), 
by the Grants-in-Aid for Scientific Research on Innovative Areas from MEXT
(Nos. 24105001, 24105008),
by the Yukawa International Program for Quark-Hadron Sciences,
and 
by HIC for FAIR and by the Polish Science Foundation (NCN),
under Maestro grant 2013/10/A/ST2/00106.


\end{document}